\newcommand{\beq}{\begin{equation}}
\newcommand{\eeq}{\end{equation}}

\newcommand{\id}
 {i\kern.06em\hbox{\raise.25ex\hbox{$/$}\kern-.60em$\partial$}}

\newcommand{\bs}{/\kern-.52em b}
\newcommand{\qs}{/\kern-.52em s}

\newcommand{\p}{\partial}
\newcommand{\yp}{^{\prime}}
\newcommand{\ts}{\tilde{e}}
\newcommand{\tp}{\tilde{\pi}}

\newcommand{\dd}
{\kern.06em\hbox{\raise.25ex\hbox{$/$}\kern-.60em$\partial$}}

\newcommand{\ep}{\epsilon}

\documentstyle[12pt]{article}
\date{}
\begin{document}
\title{SU(2) Charges as Angular-momentum in
$N=1$ Self-dual
Supergravity\footnotetext{\# Corresponding author and adress}
\thanks{Work supported in part by the National
Science Foundation
of China}}
\author{{Sze-Shiang Feng$^{\P\S\#}$,
Zi-Xing Wang$^{\dag}$
, Xi-Jun Qiu$^{\P}$}\\
$\S$ {\small {\it CCAST(World Lab.), P.O. Box. 8730, Beijing
      100080}}\\
$\P$ {\small {\it Physics Department, Shanghai
University, 201800,
Shanghai,China}}$^\#$\\e-mail:xjqiu@fudan.ac.cn$^\#$\\
$\dag$ {\small {\it Institute of Nuclear Research,
Academia Sinica,
201800, Shanghai, China}}\\
$\S$  {\small {\it Center for String Theory,
Shanghai Teacher's
   University, 200234, Shanghai, China}}}
\maketitle
\newfont{\Bbb}{msbm10 scaled\magstephalf}
\newfont{\frak}{eufm10 scaled\magstephalf}
\newfont{\sfr}{eufm7 scaled\magstephalf}
\input amssym.def
\baselineskip 0.3in
\begin{center}
\begin{minipage}{135mm}
\vskip 0.3in
\baselineskip 0.3in
\begin{center}{\bf Abstract}\end{center}
  {The $N=1$ self-dual supergravity has
  $SL(2,\Bbb C)$ symmetry
  . This symmetry
  results in $SU(2)$ charges as the angular-momentum.
As in the non-supersymmetric self-dual gravity,
the currents are also
of their potentials and are therefore identically
conserved. The charges are generally invariant and
gauge covariant under local $SU(2)$ transforms approaching
to be rigid at spatial infinity. The Poisson
brackets constitute $su(2)$ algebra and hence
can be interpreted as the generally covariant conservative
angular-momentum.
   \\PACS number(s): 11.30.Cp,11.30.Pb,04.20.Me,04.65.+e.
   }
\end{minipage}
\end{center}
\vskip 1in

\indent The study of self-dual gravities has drawn much
attention in the past decade since the discovery of
Ashtekar's
new variables, in terms of which the constraints can be
greatly
simplified\cite{s1}-\cite{s2}. The new phase variables
consist of  densitized $SU(2)$ soldering forms $\tilde{e}
^i\,_A\,^B$ from which a metric density is obtained according
to the definition $q_{ij}=-{\rm Tr}\tilde{e}_i\tilde{e}_j$,
and a complexified connection $A_{iA}\,^B$
which carries the momentum
dependence in its imaginary part.
The original Ashtekar's self-dual
canonical gravity permits also a
Lagrangian formulation\cite{s3}
\cite{s4}. The supersymmetric extension of this Lagrangian
formulation, which is equivalent to the
simple real supergravity,
was proposed by Jacobson\cite{s5}, and the corresponding
Ashtekar complex canonical transform was
given by Gorobey et al\cite{s6}.
   The Lagrangian density is\cite{s5}
\beq
{\cal L}_J=\frac{1}{\sqrt{2}}(e^{AA^{\prime}}\wedge
e_{BA^{\prime}}\wedge F_A\,^B+
ie^{AA^{\prime}}\wedge\bar{\psi}_{A^{\prime}}
\wedge{\cal D}\psi_A)
\eeq
The dynamical variables are the real tetrad
$e^{AA^{\prime}}$ (the
"real" means $\bar{e}^{A\yp A}=e^{AA\yp}$), the
traceless left-handed $SL(2.\Bbb C)$ connection
$A_{\mu MN}$
and the complex anticommuting spin-$\frac{3}{2}$
gravitino
field $\psi_{\mu A}$. The $SL(2,\Bbb C)$
covariant exterior
derivative is defined by
\beq
{\cal D}\psi_M:=d\psi_M+A_M\,^N\wedge\psi_N
\eeq
and the curvature 2-form is
\beq
F_M\,^N:=dA_M\,^N+A_M\,^P\wedge A_P\,^N
\eeq
The indices are lowered and raised with
the antisymmetric
$SL(2, \Bbb C)$ spinor $\epsilon^{AB}$ and
its inverse $\epsilon_{AB}$
according to the convention $\lambda_B=
\lambda^A\epsilon_{AB},
\lambda^A=\epsilon^{AB}\lambda_B$, and the
implied summations are always
in north-westerly fashion: from  the left-upper
to the right-lower.
The Lagrangian eq.(1) is a holomorphic functions of the
connection and the equation for $A_{\mu A}\,^B$
is equivalent
to
\beq
{\cal D}e^{AA^{\prime}}=\frac{i}{2}\psi^A\wedge
\bar{\psi}^{A^{\prime}}
\eeq
provided $e^{AA\yp}$ is real. The Lagrangian
$\frac{1}{2}({\cal L}
_J+\bar{\cal L}_J)$ for real supergravity is a
non-holomorphic function
but leads to no surfeit of field equations.
Under the left-handed
local supersymmetric transform generated by
anticommuting
parametres $\ep_A$
\beq
\delta\psi_A=2{\cal D}\ep_A,\,\,\,\,\,\,
\delta\bar{\psi}_{A\yp}=0,\,\,\,\,\,\,\,\,
\delta e_{AA\yp}=-i\bar{\psi}_{A\yp}\ep_A
\eeq
the Lagrangian ${\cal L}_J$ is invariant {\it without}
using any one
of the Euler-Lagrangian equations while
under the right-handed transform
\beq
\delta\psi_A=0,\,\,\,\,\,\,\,\,\, \delta
\bar{\psi}_{A\yp}=2{\cal D}\bar{\ep}
_{A\yp},\,\,\,\,\,\,\,\,\,\, \delta e_{AA\yp}=
-i\psi_A\bar{\ep}_{A\yp}
\eeq
${\cal L}_J$ is invariant {\it modulo}
the field equations.\\
\indent The (3+1) decomposition is effected as
\beq
{\cal L}_J=\ts^{kAB}\dot{A}_{kAB}+\tp^{kA}
\dot{\psi}_{kA}-{\cal H}
\eeq
\beq
{\cal H}:=e_{0AA\yp}{\cal H}^{AA\yp}
+\psi_{0A}{\cal S}^A
+\hat{{\cal S}}^{A\yp}\bar{\psi}_{0A\yp}
+A_{0AB}{\cal J}^{AB}
+({\rm total\,\, divergence})
\eeq
The canonical momenta are
\beq
\ts^{kAB}:=-\frac{1}{\sqrt{2}}\ep^{ijk}
e_i\,^{AA\yp}e_j^B\,_{A\yp}
\eeq
\beq
\tp^{kA}:=\frac{i}{\sqrt{2}}\ep^{ijk}e_i\,
^{AA\yp}\bar{\psi}_{jA\yp}
\eeq
and the constraints are
\beq
{\cal H}^{AA\yp}:=\frac{1}{\sqrt{2}}\ep^{ijk}(e_i\,
^{BA\yp}F_{jkB}\,^A-i\bar{\psi}_i
\,^{A\yp}{\cal D}_j\psi_k\,^A)
\eeq
\beq
{\cal S}^A:={\cal D}_k\tp^{kA}
\eeq
\beq
\hat{\cal S}^{A\yp}:=\frac{i}{\sqrt{2}}
\ep^{ijk}e_i\,^{AA\yp}{\cal D}
_j\psi_{kA}
\eeq
\beq
{\cal J}^{AB}:={\cal D}_k\ts^{kAB}-
\tp^{k(A}\psi_k\,^{B)}\
\eeq
\indent The 0-components $e_{0AA\yp},
\psi_{0A},\bar{\psi}_{0A\yp}
$ and $A_{0AB}$ are just the Lagrange
multipliers and the dynamical
conjugate pairs are $(\ts^{kAB},A_{jAB}),
(\tp^{kA}, \psi_{kA})$.
The constraints ${\cal H}^{AA\yp}=0$ and
$\hat{\cal S}^{A\yp}=0$
generate the following two
\beq
\ddot{\cal H}^{AB}:=(\ts^j\ts^kF_{jk})^{AB}
+2\tp^j\ts^k{\cal D}_{[j}\psi_{k]}\ep^{AB}
+2(\tp^j{\cal D}_{[j}\psi_{k]})\ts^{kAB}=0
\eeq
\beq
{\cal S}^{\dag A}:=\frac{1}{\sqrt{2}}
\ep^{ijk}\ts_i\,^{AB}{\cal D}
_j\psi_{kB}=0
\eeq
The equations of motion will be properly
expressed in Hamiltonian form
$\dot{f}=\{H, f\} $if we assign the Poisson brackets
\beq
\{\ts^{kAB}(x),A_{jAB}(y\}=\delta_j
\,^k\delta_{(M}\,^A\delta_{N)}
\,^B\delta^3(x,y)
\eeq
\beq
\{\tp^{kA}(x), \psi_{jA}(y)\}=-\delta_j
\,^k\delta_M\,^A\delta^3(x,y)
\eeq
all other brackets among these
quantities being zero.
This is the outline of the theory.\\
\indent  In our previous works, we have obtained
the $SU(2)$ charges
and the energy-momentum in the Ashtekar's
formulation of Einstein
gravity\cite{s7}-\cite{s8} and they are
closely related to the
angular-momentum\cite{s9}-\cite{s11} and the
energy-momentum \cite{s12}
in the vierbein formalism of Einstein gravity.
The fact that
the algebra formed by their Poisson brackets {\it do}
constitute the 3-Poincare algebra on the Cauchy surface
supports from another aspect that their
definitions are reasonable.
Similarly, the study of $SU(2)$ charges
in the self-dual supergravity
considered is also an interesting subject.
 In the following,
we will employ the $SL(2,\Bbb C)$ invariance to obtain
the conservative charges as we did previously\cite{s8}
Under any $SL(2,\Bbb C)$ transform
$$ e_{\mu AA\yp}\rightarrow L_A\,^B\bar{R}_{A\yp}\,^{B\yp}
e_{\mu BB\yp},\,\,\,\,\,\,\, \psi_A\rightarrow L_A\,^B\psi
_B,\,\,\,\,\,\,\, \bar{\psi}_{A\yp}\rightarrow
\bar{R}_{A\yp}\,^{B\yp}\bar{\psi}_{B\yp} $$
\beq
A_{\mu MN}\rightarrow L_M\,^AA_{\mu A}\,^{B}
(L^{-1})_{BN}+L_M\,^A\p_{\mu}(L^{-1})_{AN}
\eeq
${\cal L}_J$ is invariant. $L$ and $\bar{R}$ may
not neccessarily
related by complex conjugation. Note that
$L_{AB}=-(L^{-1})_{BA}$,
the transform of $A$ may also be written as
\beq
A_{\mu MN}\rightarrow L_M\,^A
L_N\,^BA_{\mu AB}-L_M\,^A\p_{\mu}L_{NA}
\eeq
For infinitesimal transform, $L_A\,^B=\delta_A\,^B+\xi_A\,^B$
where $\xi_{AB}=-\xi_{BA}$ are infinitesimal parametres. Thus
we have
\beq
\delta_{\xi}A=[\xi, A]-d\xi,\,\,\,\,\,\,\,\,
\delta \psi=\xi\psi
\eeq
When calculating the variation of the Lagrangian, one must take
into consideration of the anticommuting feature of the gravitino
field. We write the variation in the way that
\beq
\delta{\cal L}_J=\delta\phi^A(\frac{\p}{\p\phi^A}
-\p_{\mu}\frac{\p}{\p\p_{\mu}\phi^A}){\cal L}_J+\p_{\mu}
(\delta\phi^A\frac{\p}{\p\p_{\mu}\phi^A}{\cal L}_J)
\eeq
where $\phi^A$ denotes any field involved in the first order
Lagrangian. Now both $\frac{\p}{\p\phi^A}$ and
$\frac{\p}{\p\p_{\mu}\phi^A}$
are (anti-)commuting if $\phi^A$ is (anti-)commuting,
and so there is
no ordering problem.\\
\indent The invariance of ${\cal L}_J$ under the
infinitesimal $SL(2,\Bbb C)$ transform is
equivalent to the
following modulo the field equations
\beq
\p_{\rho}(\delta A_{\sigma A}\,^B
\frac{\p{\cal L}_J}{\p\p_{\rho}A_{\sigma A}\,^B}
+\delta\psi_{\sigma A}\frac{\p{\cal L}_J}
{\p\p_{\rho}\psi_{\sigma A}})=0
\eeq
For constant $\xi$, we have
\beq
\p_{\rho}(\frac{1}{\sqrt{2}}\ep^{\mu\nu\rho\sigma}
e_{\mu}\,^{AA\yp}
e_{\nu BA\yp}[\xi, A_{\sigma}]_A\,^B
+\frac{i}{\sqrt{2}}\ep^{\mu\nu\rho\sigma}e_{\mu}
\,^{AA\yp}
\bar{\psi}_{\nu A\yp}(\xi\psi_{\sigma})_A)=0
\eeq
we have therefore the conservation of $SU(2)$ charges
\beq
\p_{\mu}\tilde{j}^{\mu}_{AB}=0
\eeq
where
$$
\tilde{j}^{\rho}_{AB}=\frac{1}{\sqrt{2}}
\ep^{\mu\nu\rho\sigma}
(e_{\mu A}\,^{A\yp}e_{\nu MA\yp}A_{\sigma B}\,^M
-e_{\mu}\,^{MA\yp}e_{\nu BA\yp}A_{\sigma MA}$$
\beq
+\frac{i}{2}e_{\mu A}\,^{A\yp}
\bar{\psi}_{\nu A\yp}\psi_{\sigma B}
+\frac{i}{2}e_{\mu B}\,^{A\yp}
\bar{\psi}_{\nu A\yp}\psi_{\sigma A})
\eeq
Thus
\beq
J_{AB}=\int_{\Sigma}\tilde{j}^0_{AB} d^3x
\eeq
where
$$
\tilde{j}^0_{AB}=\frac{1}{\sqrt{2}}\ep^{ijk}
(e_{i A}\,^{A\yp}e_{j MA\yp}A_{k B}\,^M
-e_{i}\,^{MA\yp}e_{j BA\yp}A_{k MA}$$
\beq
+\frac{i}{2}e_{i A}\,^{A\yp}\bar{\psi}_{j A\yp}\psi_{kB}
+\frac{i}{2}e_{i B}\,^{A\yp}\bar{\psi}_{j A\yp}\psi_{kA})
\eeq
Using eq(9) and eq(10), $\tilde{j}^0_{AB}$ can be written as
\beq
\tilde{j}^0_{AB}=[\tilde{e}^k, A_k]_{AB}+\tp_{k(A}\psi^k_{B)}
\eeq
The constraint ${\cal J}_{AB}=0$ guarantees that
\beq
J_{AB}\approx\int_{\Sigma}\p_k\tilde{e}^k_{AB}=\int_{\p\Sigma}
\tilde{e}^k_{AB}ds_i
\eeq
where $ds_i=\frac{1}{2}\ep_{ijk}dx^j\wedge dx^k$.
It can also be obtained in the following way. Using the field
equation $e^{A\yp(A}\wedge({\cal D}e^{B)}\,_{A\yp}
-\frac{i}{2}\psi^{B)}\wedge\bar{\psi}_{A\yp})=0$, we have
$$\ep^{\rho\mu\nu\sigma}[e_{\mu A}\,^{A\yp}
(\p_{\sigma}e_{\nu BA\yp}
+A_{\sigma B}\,^Me_{\nu MA\yp}+\frac{i}{2}
\bar{\psi}_{\nu A\yp}
\psi_{\sigma B})$$
\beq
+e_{\mu B}\,^{A\yp}(\p_{\sigma}e_{\nu AA\yp}
+A_{\sigma A}\,^Me_{\nu MA\yp}+\frac{i}{2}
\bar{\psi}_{\nu A\yp}
\psi_{\sigma A})]=0
\eeq
so
\beq
\tilde{j}^{\rho}_{AB}=-\frac{1}{\sqrt{2}}
\ep^{\rho\mu\nu\sigma}
\p_{\sigma}(e_{\mu A}\,^{A\yp}e_{\nu BA\yp})
\eeq
Using
\beq
e_{[\mu A}\,^{A\yp}e_{\nu]BA\yp}=e_{[\mu AC}
e_{\nu]B}\,^C-i\sqrt{2}
n_{[\mu}e_{\nu]AB}
\eeq
we have
\beq
\begin{array}{rcl}
\tilde{j}^0_{AB}&=&-\frac{1}{\sqrt{2}}\ep^{ijk}
\p_k(e_{[iA}\,^{A\yp}e_{j]BA\yp})
=-\frac{1}{\sqrt{2}}\ep^{ijk}\p_k
(e_{[iAC}e_{j]B}\,^C-i\sqrt{2}n_{[i}e_{j]AB})\\\\
&=&\frac{1}{\sqrt{2}}\ep^{ijk}\p_k(e_ie_j)_{AB}
=\p_k\ts^{k}_{AB}
\end{array}
\eeq
which is exactly the same as eq.(30)
We can thus
have the Poisson brackets
$$
\{J_{AB}, J_{MN}\}=\{\int_{\p\Sigma}
\tilde{e}^k_{AB}ds_k,\,\,\,\,
\int_{\Sigma}(\tilde{e}^i\,_M\,^PA_{iPN}
+\tilde{e}^i\,_N\,^PA_{iPN})d^3x\}$$
\beq
=\frac{1}{2}(J_{MA}\ep_{NB}+J_{MB}\ep_{NA}
+J_{NA}\ep_{MB}
+J_{MA}\ep_{NB})
\eeq
Now the flat dreibein on $\Sigma$ is needed
in order to find the angular-
momentum $J_i$. To clarify the notions,
we use the following conventions:
$\mu,\nu,...$ denote the 4-dim curved indices
and $i,j,k,$ denote
the 3-dim curved indices on $\Sigma$; $a,b,c,.
..$ denote the flat 4-dim
indices and $ l,m,n,...$ denote the flat 3-dim
indices on $\Sigma$.
The rigid flat vierbein is denoted as $E^a_{AA\yp}$
and the rigid flat
dreibein is denoted by $E^m_{AB}$. Then define
\beq
J_m:=\frac{1}{\sqrt{2}}E_m^{AB}J_{AB}
\eeq
and using the relation $\ep^{mnl}E_mE_n
=\sqrt{2}E^l$ we have
\beq
\{J_m,J_n\}=\ep_{mnl}J^l
\eeq
 Therefore the $su(2)$ algebra is restored.
As in the non-supersymmetric case\cite{s8},
we can also obtain only
the $SU(2)$ charges instead of the whole
$SL(2,\Bbb C)$ charges.
Yet, the angualr-momentum $J_{ab}$ obtained
in\cite{s9}-\cite{s10} is
completely contained in $J_{MN}$ since we have
from eq(32) that
\beq
\tilde{j}^{\rho}_{AB}=-\frac{1}{2}\tilde{j}^{\rho}_{ab}
E^a\,_A\,^{A\yp}E^b\,_{BA\yp}
\eeq
where $\tilde{j}^{\rho}_{ab}$ is the angular
-momentum current
obtained in \cite{s9}-\cite{s10}.
\beq
\tilde{j}^{\rho}_{ab}=\sqrt{2}\ep^{\rho\sigma\mu\nu}
\p_{\sigma}
(e_{\mu a}e_{\nu b})
\eeq
and the angualr-momentum is
\beq
J_{ab}=\int_{\Sigma}\tilde{j}^0_{ab} d^3x
\eeq
Hence
\beq
\begin{array}{rcl}
J_{MN}&=&-\frac{1}{2}J^{ab}E_{[aM}\,^{A\yp}E_{b]NA\yp}=
-\frac{1}{2}(J^{ij}E_{[iAC}E_{j]}\,^{BC}-i\sqrt{2}
J^{0i}n_0E_{iA}\,^B)\\\\
&=&\frac{1}{\sqrt{2}}(L_i-iK_i)E^i_{MN}
\end{array}
\eeq
where $L_i=\frac{1}{2}\ep_{ijk}J^{jk}$ are the
spatial rotations
and $K_i=J_{0i}=-J^{0i}$ are the Lorentz boosts.
Therefore
\beq
J_i=\frac{1}{2}(L_i-iK_i)
\eeq
Bear in mind that both $\frac{1}{2}(L_i-iK_i)$
and $\frac{1}{2}(L_i+iK_i)$ obey the $su(2)$
algbra\cite{s13}. Actually, the boost
charges are vanishing as can be seen
from eq(30).
Thus we can obtain the angualr-momentum,
 in the self-dual simple supergravity once
 $J_{MN}$ is known.\\
\indent We make a few remarks finally.
 The total charges take the same
 integral form as those in the
non-supersymmetric case. Though we can
obtain the $SU(2)$
sector of the $SL(2,\Bbb C)$ charges, the
information
of the angular-momentum is completely
contained in the
$SU(2)$ charges. It can be seen from the
surface integrals
that the angular-momentum
is governed by the $r^{-2}$ part of $\ts^i$.
 As in \cite{s1}-\cite{s2}, we always
 assume that
the phase space variables are subject
to the boundary
conditions.
\beq
e^{\mu}_{AB\mid\p\Sigma}
=(1+\frac{M(\theta,\phi)}{r})^2\stackrel{0}{e
^{\mu}}_{AB}+O(1/r^2),\,\,\,\,
A_{\mu MN}\,_{\mid\p\Sigma}=O(1/r^2)
\eeq
\beq
\tp^i_A=O(1/r),\,\,\,\, \psi_{\mu A}=O(1/r)
\eeq
where $\stackrel{0}{e^{\mu}}_{AB}$ denote the
flat $SU(2)$ soldering forms.
As a consequence, under the $SL(2,\Bbb C)$
transforms behaving as
\beq
L_A\,^B=\Lambda_A\,^B+O(1/r^{1+\ep}), \,\,\,\,\,\,(\ep>0)
\eeq
where $\Lambda$ are rigid transforms
The charges transform as
\beq
J_{MN}\rightarrow\Lambda_M\,^A\Lambda_N\,^BJ_{AB},\,\,\,
\eeq
i.e., they gauge covariant.
Their conservation is generally covariant. As in the
non-supersymmetric case\cite{s7}-\cite{s8}
, the currents have also potentials, i.e., can
be expressed as a divergence of a antisymmetric
tensor density. So they are identically conserved.
Upon quantization,
the Poisson brackets correspond to the quantal commutators
 and their algebra
realizes indeed the $su(2)$ algebra. This shows that
their interpertations reasonable.\\
\indent It is novel that the relation between $J_{MN}$
and the constraint
${\cal J}^{AB}$ is the same as that between
the electric charge
and the Gauss law constraint in QED\cite{s14}
\beq
{\bf \nabla}\bullet{\bf E}-e\bar{\psi}\gamma_0\psi=0
\eeq
\beq
q=\int_{\p\Sigma} {\bf E}\bullet d{\bf S}
\eeq
So the $J_{MN}$ is a kind of gauge charge.

\end{document}